\documentclass[ pre, preprint, superscriptaddress, longbibliography, nobalancelastpage]{revtex4-2}
\usepackage{graphicx}
\usepackage{tabularx}
\usepackage{amssymb}
\usepackage{amsmath}
\usepackage{hyperref}
\usepackage{float}
\usepackage{mathtools}
\usepackage{bm}
\usepackage{dcolumn}
\setcitestyle{super}

\begin{document}

\title{A minimal coarse-grained model to study the gelation of multi-armed DNA nanostars}

\author{Supriyo Naskar}
\email{supriyo@iisc.ac.in}
\affiliation{Center for Condensed Matter Theory, Department of Physics, Indian Institute of Science, Bangalore 560012, India}
\author{Dhiraj Bhatia}
\email{dhiraj.bhatia@iitgn.ac.in}
\affiliation{Center for Biomedical Engineering, Indian Institute of Technology Gandhinagar, Palaj, Gandhinagar, Gujarat 382355, India}

\author{ Shiang-Tai Lin}
\email{stlin@ntu.edu.tw}
\affiliation{Department of Chemical Engineering, National Taiwan University, Taipei 10617, Taiwan}

\author{Prabal K. Maiti}
\thanks{Corresponding author}
\email{maiti@iisc.ac.in}
\affiliation{Center for Condensed Matter Theory, Department of Physics, Indian Institute of Science, Bangalore 560012, India}
\date{\today}
\begin{abstract}
DNA is an astonishing material that can be used as a molecular building block to construct periodic arrays and devices with nanoscale accuracy and precision. Here, we present simple bead-spring model of DNA nanostars having three, four and five arms and study their self-assembly using molecular dynamics simulations. Our simulations show that the DNA nanostars form thermodynamically stable fully bonded gel phase from an unstructured liquid phase with the lowering of temperature. We characterize the phase transition by calculating several structural features such as radial distribution function and structure factor. The thermodynamics of gelation is quantified by the potential energy and translational pair-entropy of the system. The phase transition from the arrested gel phase to an unstructured liquid phase has been modelled using two-state theoretical model. We find that this transition is enthalpic driven and loss of configuration and translational entropy is counterpoised by enthalpic interaction of the DNA sticky-ends which is giving rise to gel phase at low temperature. The absolute rotational and translational entropy of the systems, measured using two-phase thermodynamic model, also substantiate the gel transition. The slowing down of the dynamics upon approaching the transition temperature from a high temperature, demonstrating the phase transition to the gel phase.  The detailed numerical simulation study of the morphology, dynamics and thermodynamics of DNA gelation can provide guidance for future experiments, easily extensible to other polymeric systems, and has a remarkable implications in the DNA nanotechnology field.
\end{abstract}

\maketitle
\newpage

\section{\NoCaseChange{Introduction}}
The unique base-pairing specificity rule of DNA makes it an ideal building block to design highly organized supramolecular materials with nanoscale accuracy\cite{chap7_1}. Purposely sequenced single-strand DNA bases can form
DNA motifs via reciprocal exchange between adjacent duplexes. These motifs via hierarchical self-assembly, form a wide range of highly organized artificial nanostructures. The concept of DNA nanotechnology has it's 
root way back in 1980s when  Ned Seeman of New York University first designed an immobile four way DNA junction using four different single-stranded DNA\cite{chap7_SEEMAN1982237,chap7_kallenbach1983immobile}. This idea led the foundation of structural DNA nanotechnology field to construct synthetic two-dimensional and even three-dimensional DNA nanostructures\cite{chap7_seeman2003dna,chap7_winfree1998design}. Later, the ``origami" method revolutionized the field by enabling creation of more complex and larger DNA nanostructures\cite{chap7_rothemund2006folding}. Since then the field expanded rapidly due to its potential applications in diverse fields such as nanomedicine, nanoelectronics, synthetic biology, nanorobotics\cite{chap7_7,chap7_30,chap7_31,chap7_59,chap7_69,chap7_bhatia2011synthetic,chap7_dna_nanoelec}.\par

DNA can be assembled into complex 1D, 2D and 3D DNA nanostructures with precise  arrangements of helices either 
via DNA origami or DNA tile techniques. In similar fashions,  small DNA nanostructures can self-assemble into the complex polymeric network at desired temperature forming 
hydrogels\cite{chap7_luo1,chap7_lee2012mechanical,chap7_cheng2009ph,chap7_dhiraj_gel,chap7_reviewgel}. The resulting gel formed via hydrogen bonding  and other dispersive interactions can be classified as physical hydrogels and are biodegradable, bio-compatible, nontoxic. They can encapsulate drugs, cargo, and even cells that have extensive applications such as controlled drug delivery, 3D cell culture, bio-printing, cell transplant therapy, tissue engineering and other biomedical applications\cite{chap7_luo1,chap7_lee2012mechanical,chap7_cheng2009ph,chap7_dhiraj_gel,dhiraj}. The mechanism of DNA hydrogel formation is rather simple. Single-stranded DNA first hybridized to form small DNA nanostars with sticky ends and then those nanostars exploiting the sticky-end cohesion, bind with each other to form a spanning polymeric network. By tuning the length and sequence of sticky-ends, it is even possible to thermally control the gel formation. Nagahara and Matsuda first reported of hybrid polyacrylamide-DNA hydrogel\cite{chap7_NAGAHARA1996111}. Later Liu, Zhou, and co-workers using a one-pot step-wise self-assembly protocol prepared the first pure DNA supramolecular hydrogel\cite{chap7_liu2009}.  In 2006, Luo et al. fabricated a hydrogel under physiological conditions entirely from branched DNA by enzymatic ligation which exhibited amazing shape memory properties\cite{chap7_luo1}. Several other experimental observation has been reported on different aspects and applications of DNA hydrogel\cite{chap7_expt1,chap7_expt2,chap7_expt3,chap7_expt4,chap7_expt5,chap7_expt6,chap7_expt7,chap7_expt8,chap7_expt9,chap7_expt10,chap7_expt11,dhiraj}. More recently, DNA based hydrogels is utilized to stimulate membrane endocytosis, which leads to enhanced cell spreading and invasion for cells\cite{dhiraj}. Notably, Eiser et al. extensively probed the microrheological and viscoelastic properties of DNA hydrogels\cite{chap7_Xing8137}. The collective phase behaviour and dynamics of DNA hydrogels have been investigated both experimentally and numerically\cite{chap7_expt9,chap7_sim1,chap7_sim2,chap7_sim3,chap7_sim4,chap7_sim5,chap7_sim6,chap7_sim7,chap7_sim8,chap7_sim9,chap7_largo2007self,chap7_sim10}. \par

One of the critical issues of DNA hydrogel paradigm is predicting the complex three-dimensional self-assembled structures. The  kinetics of the multidimensional aggregation process are relevant to answer many fundamental physics issues. Also, how the morphology of constituent DNA nanostars effect the disordered arrested gel states is known poorly. Most of the recent studies focus mainly on the synthesis, modelling and applications of the DNA hydrogel and the fundamental physics of gelation still lacks a good understanding. Hence a detailed study answering these issues is the 
need of the moment. \par
All-atom models which was found to be very effective in addressing self-assembly and mechanical properties of DNA nanostructures are too expensive to study DNA hydrogel formations\cite{chap7_all3,chap7_all2,chap7_all1,chap7_all4}. Coarse-grained (CG)  model with different levels of descriptions have found to be more effective in exploring different aspects of the gelation\cite{chap7_sim2,chap7_sim3,sn2,chap7_sim4,chap7_sim5,chap7_sim6,chap7_largo2007self}. Although, most of the base-pair level coarse-grained models are too detailed and computationally demanding -- thus often not useful to study the self-assembly of DNA nanostars\cite{chap7_cg1,chap7_cg2,chap7_cg3,chap7_cg4,chap7_cg5,chap7_cg6}. Recently, a simple bead-spring model has been introduced by Xing et al. in which the sticky ends are replaced by single patch\cite{chap7_xing2019}. The model has captured the self-assembly and microrheological properties of gelation of three armed ``Y" shaped DNA nanostars. \par
In this work, using the simple bead-spring model we have studied the structure and thermodynamics of gelation of DNA nanostars having more complex morphologies. We have modelled three different classes of DNA nanostar: (i) trivalent nanostar (referred to as Y-DNA), (ii) tetravalent nanostar (referred to as X-DNA) and (iii) pentavalent nanostar (referred to as 5WJ). In the methodology section, a brief description of models and methods implemented in this study is given. In the result sections, we first discussed the structure of DNA hydrogel, followed by thermodynamics of gelation where we explicitly evaluated the potential energy, translational pair entropy and change in enthalpy and entropy upon gelation. Then, we employed two-phase thermodynamic method (2PT) to calculate the absolute translational and orientational entropy to distinguish different phases of hydrogel. Finally, the dynamics is probed by velocity autocorrelation and mean square displacement of the systems.  
\newpage
\section{\NoCaseChange{Models and Numerical Methods}}
\subsection{\NoCaseChange{Coarse-grained model}}
\begin{figure*}[htbp]
 \centering
 \includegraphics[width=\linewidth,keepaspectratio=true]{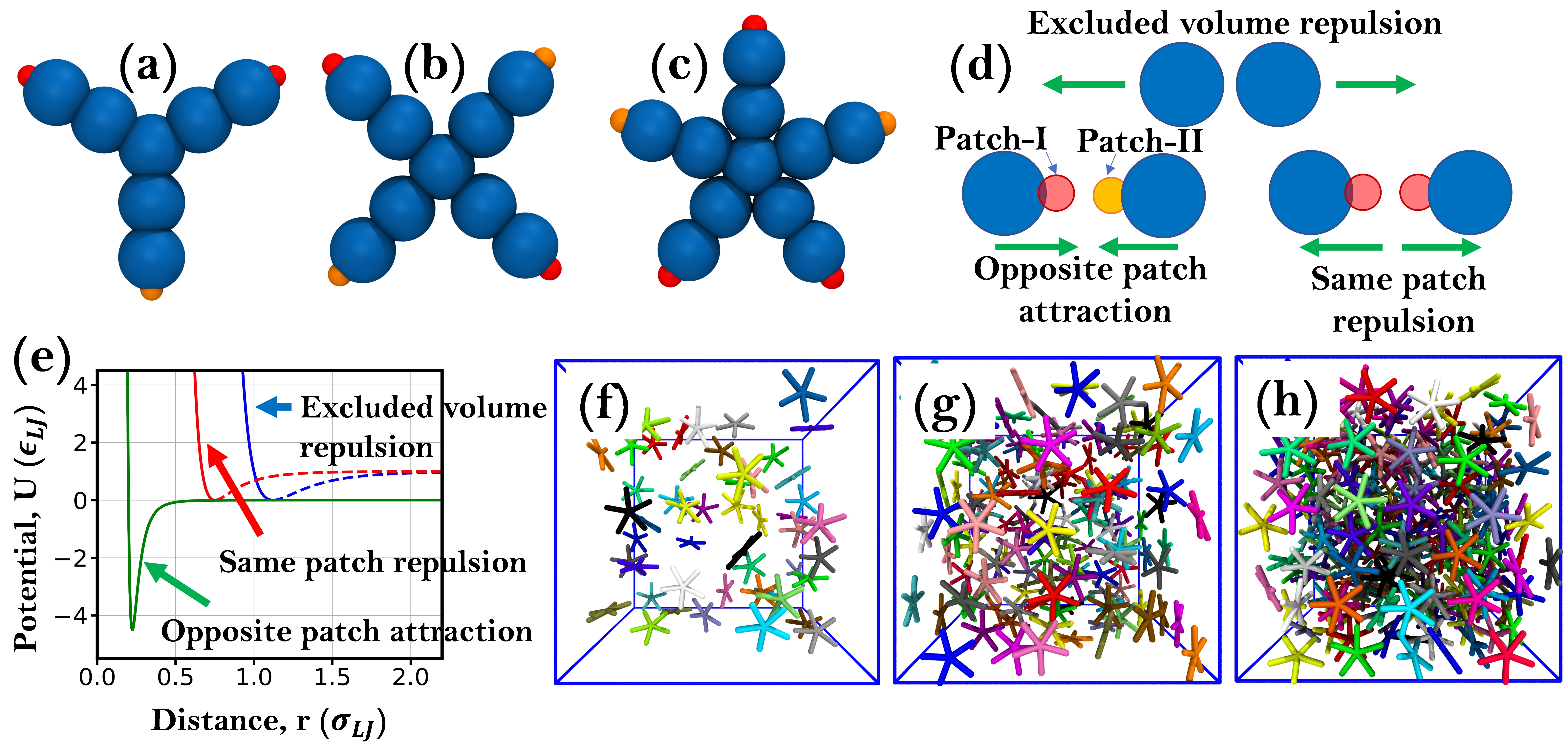}
 \caption{Coarse-grained models of the DNA nanostars investigated in this study. (a) Y-shaped DNA nanostar, (b) X-shaped DNA nanostar and (c) 5 armed (5WJ) DNA nanostar. (d) Non-bonded interactions between different CG beads. (e) Pairwise LJ potential between different beads. (f)-(h) Initial configuration of the system with different densities. The dimension of each box is 30$\times$30$\times$30 $\sigma_{LJ}^{3}$. The number of nanostars in each simulation box are (f) $N=50$, (g) $N=150$, (h) $N=250$ respectively.}
 \label{ch7_fig1}
\end{figure*}

We employ a bead-spring CG model to represent the structure of DNA nanostars which was originally developed by Eiser and coworkers\cite{chap7_xing2019}. All the force field (FF) parameters like mass ($m_{LJ}$), length ($\sigma_{LJ}$), energy ($\epsilon_{LJ}$), and the Boltzmann constant ($k_B$)  are expressed in reduced Lennard-Jones (LJ) unit. In terms of these quantities, the reduced time is defined as $\tau_{LJ} = \sqrt{m\sigma^{2}_{LJ}/\epsilon_{LJ}} $. The CG structure is composed of two types of charge-less beads -- one large bead to replicate the double-stranded nature of DNA and one small patch to represent the sticky ends [figure \ref{ch7_fig1}]. Two types of small patch are defined (designated by patch-I and patch-II). An attractive potential is implemented between the opposite patches only while a repulsive potential is applied between same type of patches. This is implemented to prevent any three-body interaction between the patches. The non-bonded excluded volume interaction between the beads are modelled using truncated and shifted LJ potential, known as Weeks-Chandler-Andersen (WCA) potential truncated at the minima of the potential (distance $1.12\sigma_{LJ}$). 
\begin{equation}
U_{WCA}(r,\sigma,\epsilon) =
\begin{cases} 
 4\epsilon \left[\left(\dfrac{\sigma}{r}\right)^{12}-\left(\dfrac{\sigma}{r}\right)^{6}\right] + U^{\prime} &\text{$ \forall\ r\le 1.12\sigma_{LJ}$}
\\
 0  &\text{$\forall\ r\ge 1.12\sigma_{LJ}$} 
 \label{ch7_1}
 \end{cases}
\end{equation}
Where, r is the distance between two particles, $\epsilon$ is the depth of the potential $\sigma$ is sigma is the size of the particle. $ U^{\prime}$ is set in such a way that $U_{WCA}(r=1.12\sigma_{LJ})=0$. In our simulation, we set $\epsilon = \epsilon_{LJ}$, $\sigma = \sigma_{LJ}$ and $r_{cutoff} = 1.12\sigma_{LJ}$ for the interaction between two large beads. This leads to a repulsive potential which prevents any unwanted overlap between the arms of the nanostars. For non-complementary patches, we set, $\epsilon = \epsilon_{LJ}$, $\sigma = 0.67\sigma_{LJ}$ and $r_{cutoff} = 0.67\sigma_{LJ}$ which generates a repulsive interaction between the same patches and prevent multiple attraction between the different patchy ends. Suppose a bond between patch-I and patch-II is formed, then the repulsive potential prevents any possible bond formation of patch-I--patch-I or patch-II -- patch-II which gives each arm a strict valency of 1 [figure \ref{ch7_fig1} (d)]. The attractive interaction between the complementary patches are described by full LJ potential truncated at a longer distance as given by equation \ref{ch7_lj},
\begin{equation}
U_{LJ}(r,\sigma,\epsilon) =
\begin{cases} 
 4\epsilon \left[\left(\dfrac{\sigma}{r}\right)^{12}-\left(\dfrac{\sigma}{r}\right)^{6}\right]  &\text{$ \forall\ r\le 5\sigma_{LJ}$}
\\
 0  &\text{$\forall\ r\ge 5\sigma_{LJ}$} 
 \label{ch7_lj}
 \end{cases}
\end{equation}
For patch-I and patch-II interaction, we set $\epsilon = 4.5\epsilon_{LJ}$ and $\sigma = 0.2\sigma_{LJ}$. This leads to an attractive potential at a very short distance [figure \ref{ch7_fig1} (e)], which mimics the sticky-end attraction of different DNA arms. \par

All the neighbouring beads are connected by a harmonic potential of the form, 

\begin{equation}
U_{bond} = K_r (r-r_0)^2 \label{ch7_2}
\end{equation}

Where, $K_r$ is the spring constant and $r_0$ is the equilibrium bond length. The value of $K_r$ is set to a very high value of $300\epsilon_{LJ}/\sigma_{LJ}^2$ to allow only small fluctuation in the bond length and maintain the structural integrity of the nanostars. The value of $r_0$ between two large bead is set to $0.96\sigma_{LJ}$ and between large bead and any patchy particle is set to $0.56\sigma_{LJ}$. \par
Similarly, the angle bending potential involving three beads is given by the following harmonic potential,

\begin{equation}
U_{angle} = K_{\theta} ({\theta}-{\theta}_0)^2 \label{ch7_3}
\end{equation}

Where, $K_{\theta}$ is the spring constant and ${\theta}_0$ is the equilibrium bond angle. Like $K_{r}$, the value of $K_{\theta}$ is also set to a very high value of $300\epsilon_{LJ}/radian^2$ to maintain the Y-shape geometry. The equilibrium bond angles involving central bead ${\theta}_0$ = 120$^o$ for Y-shaped nanostars, 90$^o$ for X-shaped nanostars, and 72$^o$ 5WJ nanostars. The equilibrium bond angle involving three beads in the branch of the nanostar is 180$^o$. \par

\subsection{\NoCaseChange{System and simulation details}}
The schematics of the coarse-grained model of the DNA nanostars (Y-shaped, X-shaped and 5WJ) are shown in figure \ref{ch7_fig1}. For each of the DNA nanostars, we built three systems having 50, 150 and 250 units. We place the nanostars in a periodic box with random position and orientation. The dimension of the of periodic box is 30$\times$30$\times$30 $\sigma_{LJ}^{3}$. The density, volume fraction and other details of the simulated systems are given in table \ref{ch7_table1}. The volume fraction is calculated based on the volume of each nanostar. The volume of the each arm is around  $\sim \pi r^{2} l$ with $r = 0.56\ \sigma_{LJ}$ and $l = 2.48\ \sigma_{LJ}$. Based on this calculation, the volume of one Y-shaped, X-shaped and 5WJ nanostar is $7.33\ \sigma_{LJ}^{3}$, $9.77\ \sigma_{LJ}^{3}$,  $12.22\ \sigma_{LJ}^{3}$ respectively.
\begin{table}[htbp]
\centering
\caption{Details of the studied systems.}
\label{ch7_table1}
\newcolumntype{Y}{>{\centering\arraybackslash}X}
\begin{tabularx}{\textwidth}{@{}YYYY@{}}
\\
\hline 
\hline
System & Number of nanostars & Number Density($\times 10^{-3}$) & Volume Fraction (\%) \\
\hline
Y-DNA  & 50                  & 1.85                             & 1.35                 \\
       & 150                 & 5.55                             & 4.05                 \\
       & 250                 & 9.25                             & 6.75                 \\
\hline
X-DNA  & 50                  & 1.85                             & 1.81                 \\
       & 150                 & 5.55                             & 5.43                 \\
       & 250                 & 9.25                             & 9.05                 \\
\hline
5WJ    & 50                  & 1.85                             & 2.26                 \\
       & 150                 & 5.55                             & 6.78                 \\
       & 250                 & 9.25                             & 11.30                \\
\hline
\hline
\end{tabularx}
\end{table}
\par

We performed CG molecular dynamics simulation using Lammps simulation package. The time evolution of the position of each bead was done by employing Langevin equation of motion given by,
\begin{equation}
m\cfrac{d^{2}\boldsymbol{r}}{dt^{2}}= -\nabla U_{tot} - \gamma \cfrac{d\boldsymbol{r}}{dt} + \xi (t) \label{ch7_4}
\end{equation}
Where $\boldsymbol{r}$ and $m$ are the position and mass of a bead respectively. $U_{tot}$ is the total potential energy given as $U_{tot} = U_{WCA}+U_{bond}+U_{angle}$. $\gamma$ is the damping factor. The value of $\gamma$ is set to a high value of 100 to mimic the over-damped condition of gel phase. $\xi (t)$ is the stochastic noise coming from the interaction of particle with the heat bath, which can be written as, $\xi (t) = \sqrt{2 \gamma k_B T} R ( t )$. $R(t)$ is delta correlated Gaussian white noise. For each system, we varied the temperature from $ 0.075\ \epsilon_{LJ} / k_{B} $ to $ 0.775\ \epsilon_{LJ} / k_{B}$ with an interval of $0.025\ \epsilon_{LJ} / k_{B}$. For more details of the model and simulation details readers are refereed to the original paper by Eiser et al.\cite{chap7_xing2019}. The time step of integration is set to 0.005 $\tau_{LJ}$. For each temperature, we run a total of $10^{7}$ steps. To ensure equilibration, we make sure that the energy and the number of bonds formed attains a stationary value. 

\begin{figure*}[htbp]
 \centering
 \includegraphics[width=\linewidth,keepaspectratio=true]{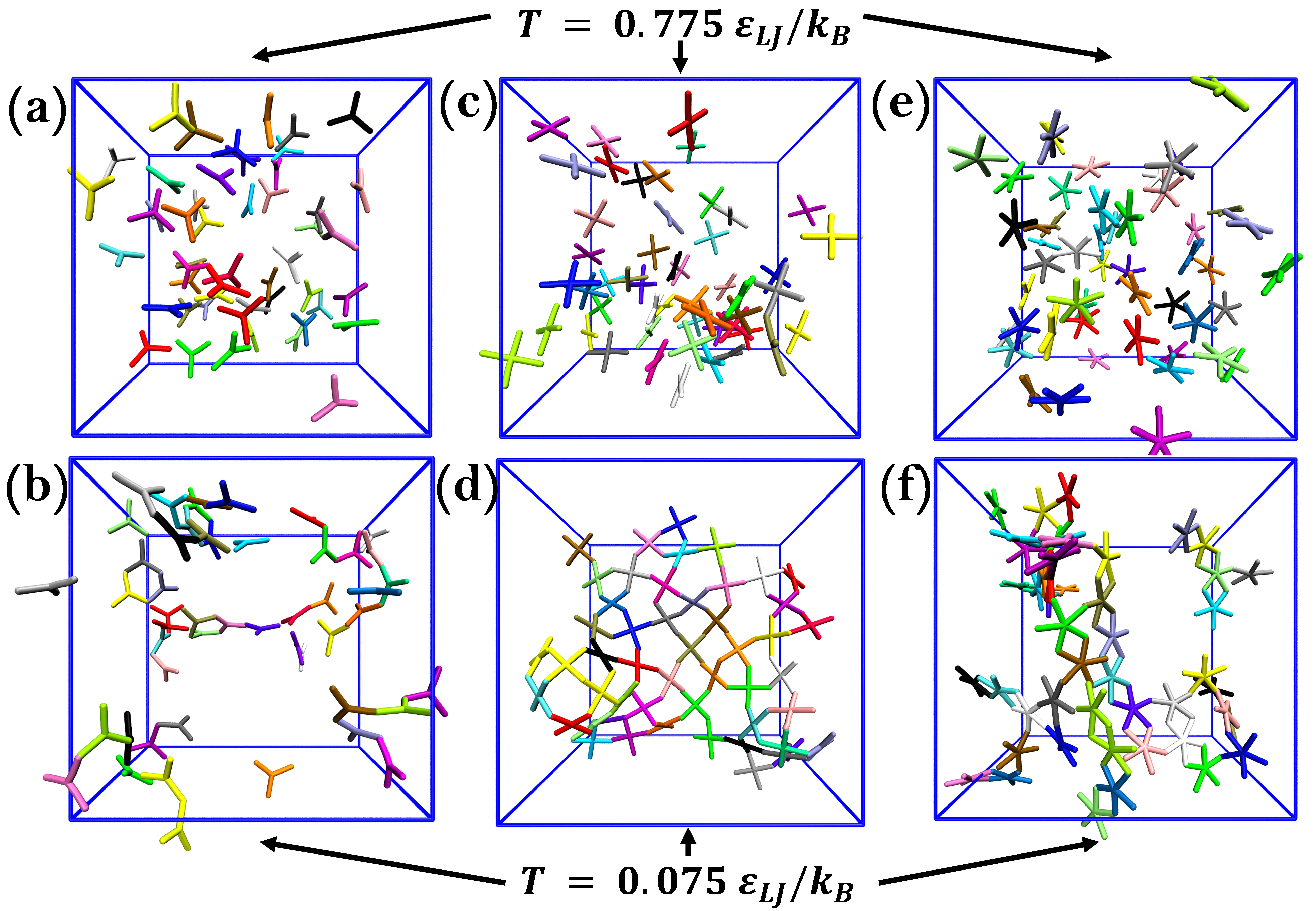}
 \caption{Structure of the DNA nanostar mixture at different temperatures. Y-DNA at temperature (a) $ T = 0.775\ \epsilon_{LJ} / k_{B}$ and (b) $T = 0.075\ \epsilon_{LJ} / k_{B}$. X-DNA at temperature (c) $ T = 0.775\ \epsilon_{LJ} / k_{B}$ and (d) $T = 0.075\ \epsilon_{LJ} / k_{B}$. 5WJ DNA at temperature (e) $ T = 0.775\ \epsilon_{LJ} / k_{B}$ and (f) $T = 0.075\ \epsilon_{LJ} / k_{B}$. The number of nanostars in each system is, $N=50$.}
 \label{ch7_fig2}
\end{figure*}

\section{\NoCaseChange{Results and Discussion}}
\subsection{\NoCaseChange{Structure}} 
In order to see the formation the gel phase, we have explored a wide range of temperatures and densities. At high temperature ($T^{*}>0.5$), the nanostars are randomly oriented in a gas like structure. As the temperature is decreased, the nanostars self-assemble into complex percolating network [figure \ref{ch7_fig2}]. In most cases such network percolates over the whole simulation box. To characterize the structure of such complex networks, we have calculated the radial distribution function (RDF), 
\begin{equation}
g(r) = \cfrac{1}{4\pi r^{2} \rho N} \sum_{i=1}^{N} \sum_{j \neq i}^{N} \langle \delta |r_{ij}-r| \rangle \label{ch7_5}
\end{equation}
\begin{figure}[htbp]
 \centering
 \includegraphics[width=\linewidth,keepaspectratio=true]{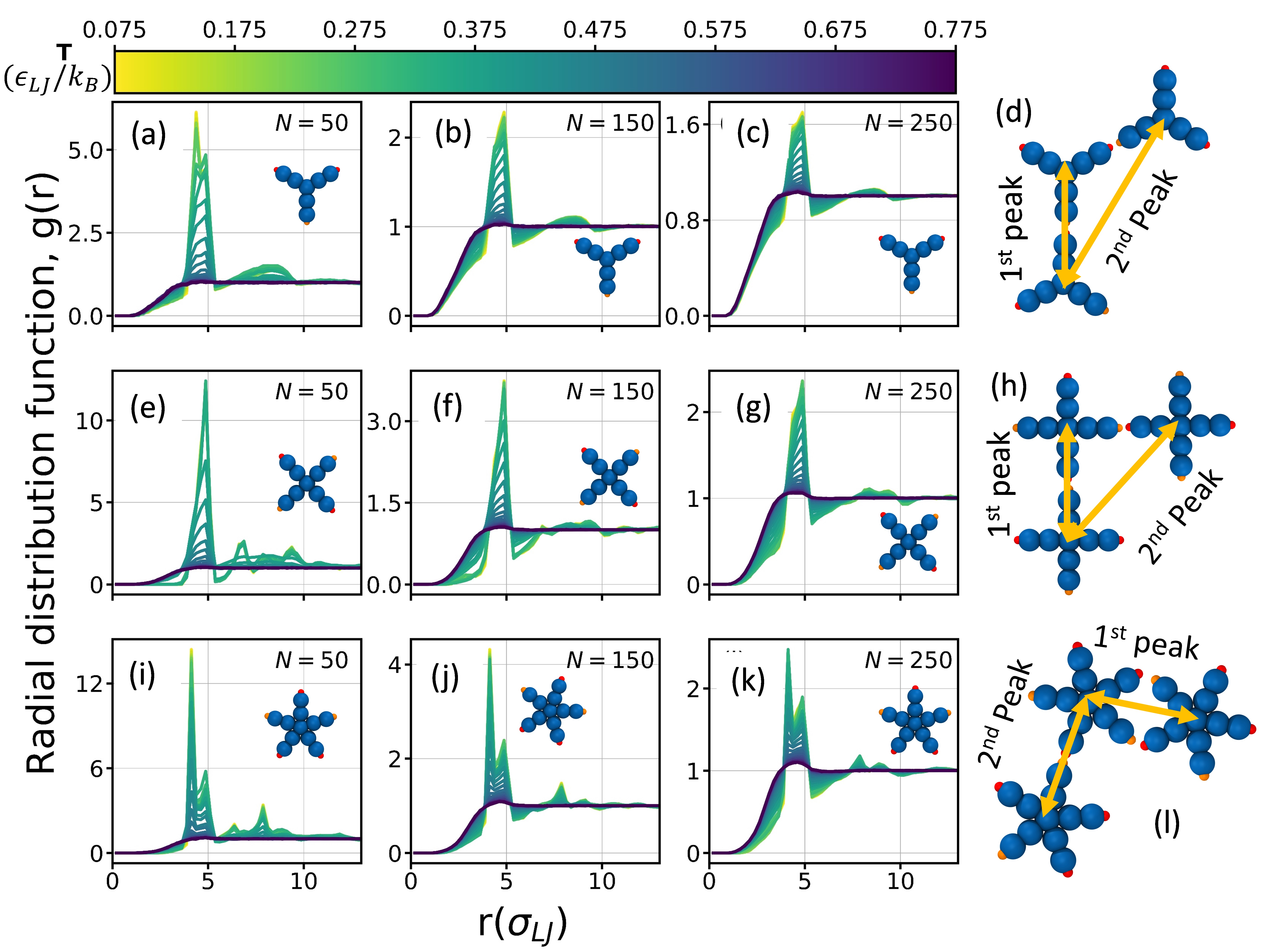}
 \caption{Radial distribution function (RDF) of the nanostar mixtures at different temperature and densities. (a)-(d) RDF of Y-shaped nanostar mixtures. The number of nanostars in the systems is (a) $N=50$, (b) $N=150$, (c) $N=250$. (d) Location of the 1st and 2nd peak in the RDF when two Y-shaped nanostars are bonded. (e)-(h) RDF of X-shaped nanostar mixtures. The number of nanostars in the systems is (e) $N=50$, (f) $N=150$, (g) $N=250$. (h) Location of the 1st and 2nd peak in the RDF when two X-shaped nanostars are bonded. (i)-(l) RDF of 5WJ nanostar mixtures. The number of nanostars in the systems is (i) $N=50$, (j) $N=150$, (k) $N=250$. (l) Location of the 1st and 2nd peak in the RDF when two 5WJ nanostars are bonded. }
 \label{ch7_fig3}
\end{figure}
Where $N$ is the total number of nanostars, $\rho$ is the averaged number density of systems, and $\langle ... \rangle$ denotes ensemble average. The sum counts the total number of pairs at the distance $r$. The RDF of the central bead of each nanostar is calculated using VMD software\cite{chap7_vmd} and has been shown in figure \ref{ch7_fig3}. At high temperature ($T^{*}>0.5$), the nanostars form a gas like unstructured fluid and the RDF is almost flat with no peaks. With the lowering of temperature ($T^{*}<0.3$),  the nanostars form networks and we see the emergence of clear peaks in the RDF. The position of the first peak in the RDF can be estimated from geometric criteria of the  bonded nanostars. The length of an arm of a nanostar is around $2.48\ \sigma_{LJ}$. When two nanostars are bonded, the distance between their central beads  is around $4.96\ \sigma_{LJ}$. The location of the first peak of the RDF is also at a separation of $4.96\ \sigma_{LJ}$. Similarly, the location of the 2nd and 3rd peaks can also be described by using different arrangements of nanostars in the network [figure \ref{ch7_fig3} (d), (h), (l)] Also with increasing densities,  we observe a decrease in the value of the peak height of the first peak of the RDF. As the density increases the number of available pairs between  $r$ to $r+dr$ remains almost same but the total number of nanostars, $N_{pairs}$ and number density, $\rho$ increases. As a result, the height of the first peak of the $g(r)$ decreases. \par

\begin{figure}[htbp]
 \centering
 \includegraphics[width=0.79\linewidth,keepaspectratio=true]{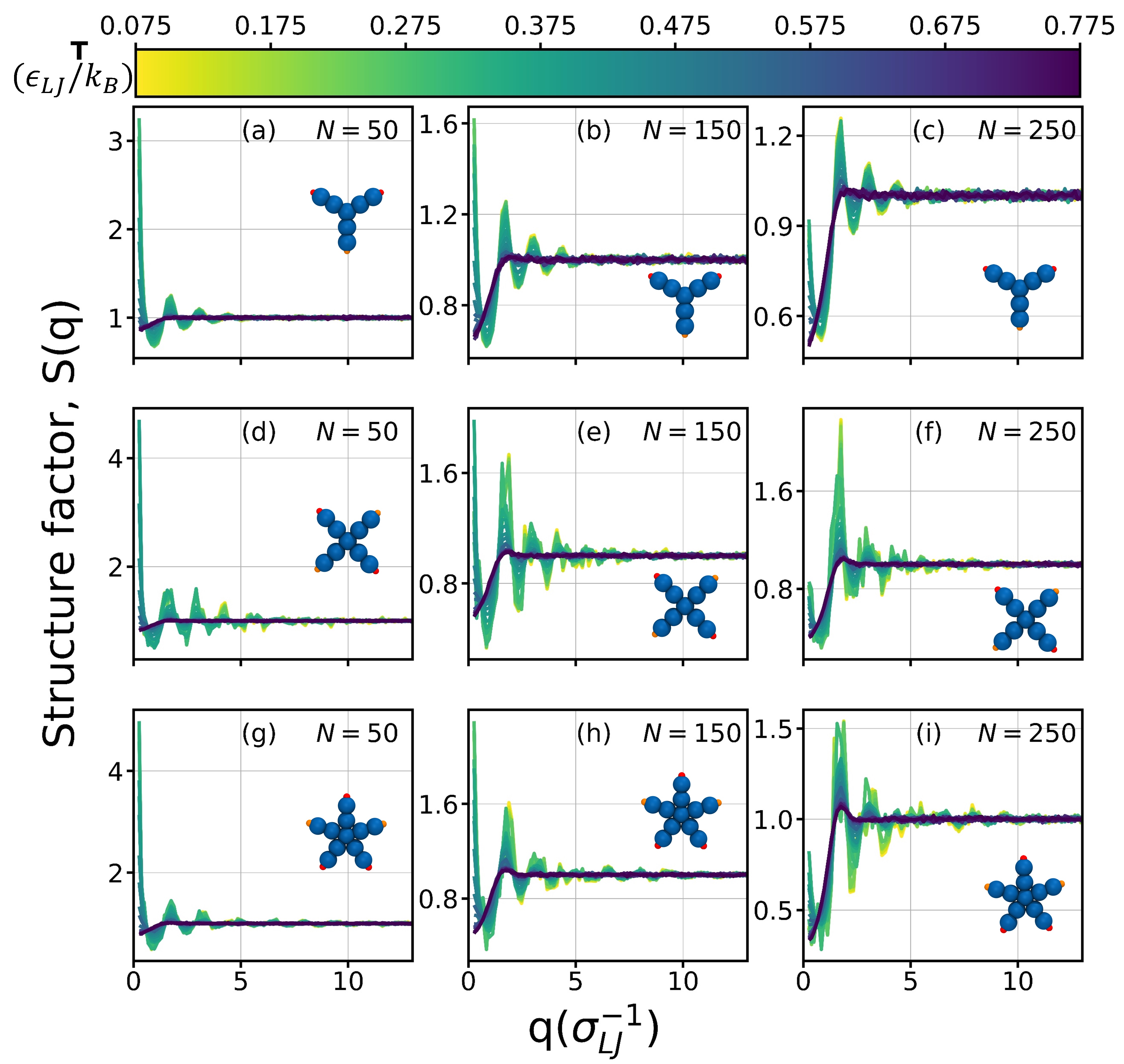}
 \caption{Structure factor (SF) of the nanostar mixtures at different temperature and densities. (a)-(c) SF of Y-shaped nanostar mixtures. The number of nanostars in the systems is (a) $N=50$, (b) $N=150$, (c) $N=250$. (d)-(f) SF of X-shaped nanostar mixtures. The number of nanostars in the systems is (d) $N=50$, (e) $N=150$, (f) $N=250$. (g)-(i) SF of 5WJ nanostar mixtures. The number of nanostars in the systems is (g) $N=50$, (h) $N=150$, (i) $N=250$.}
 \label{ch7_fig4}
\end{figure}

Another interesting fact is that the network structure is amorphous\cite{chap7_expt9}. However, from the RDF calculation, it is not clear whether the structures show crystalline or noncrystalline phase. To justify the absence of the crystalline ordering in the network, we have computed the structure factor (SF), $SF(q)$. The SF of central beads has been calculated by taking the Fourier transform of the RDF, 
\begin{equation}
SF(q) = \cfrac{1}{N} \langle \sum_{i=1}^{N} \sum_{j \neq i}^{N} exp \left[i\Vec{q}\cdot(\Vec{r_i}-\Vec{r_j}) \right]  \rangle    \label{ch7_6}
\end{equation}
Where $q$ is the wave vector, $r_i$ is the position of the particle, $\langle ... \rangle$ denotes ensemble average. The structure factors of the central bead of various nanostar mixtures are shown in figure \ref{ch7_fig4}. Similar to RDF,  $SF(q)$ shows flat profile indicating that mixtures form a unstructured fluid at high temperature ($T^{*}>0.5$). At low temperature ($T^{*}<0.3$), DNA nanostars undergo structural ordering. However, we do not see any signature of crystallization. The absence of any Brag's peak in the $SF(q)$ substantiate the amorphous nature of the network. Indeed, in several previous experimental and theoretical studies, it has been reported that the gels of DNA never crystallize\cite{chap7_expt9,chap7_largo2007self}. Our calculation of $SF(q)$ reaffirms the observation. At very low T, we observe a trihedral network for Y-DNA mixtures (observed by number of peaks in $SF(q)$), whereas for X-DNA we see tetrahedral order. For 5WJ DNA nanostars,  we find a pentagonal order at very low T. At high densities, i.e. when the number of nanostars in the system is $N=250$, the trihedral/tetrahedral/pentahedral network of the different nanostar systems becomes more prominent. Also, at lower densities we observe an increase  in $SF(q)$ as $q \rightarrow 0$. This is an indication of a high compressibility at low density. At low density, liquid like characteristics are more dominant. As a result, we find high compressibility. However, as we increase the nanostar density gel characteristics starts to dominate and the compressibility reduces. Similar observations was reported by Sciortino et al. in their study of tetra-valent DNA hydrogel\cite{chap7_largo2007self}.

\subsection{\NoCaseChange{Thermodynamics}}
\begin{figure}[htbp]
 \centering
 \includegraphics[width=\linewidth,keepaspectratio=true]{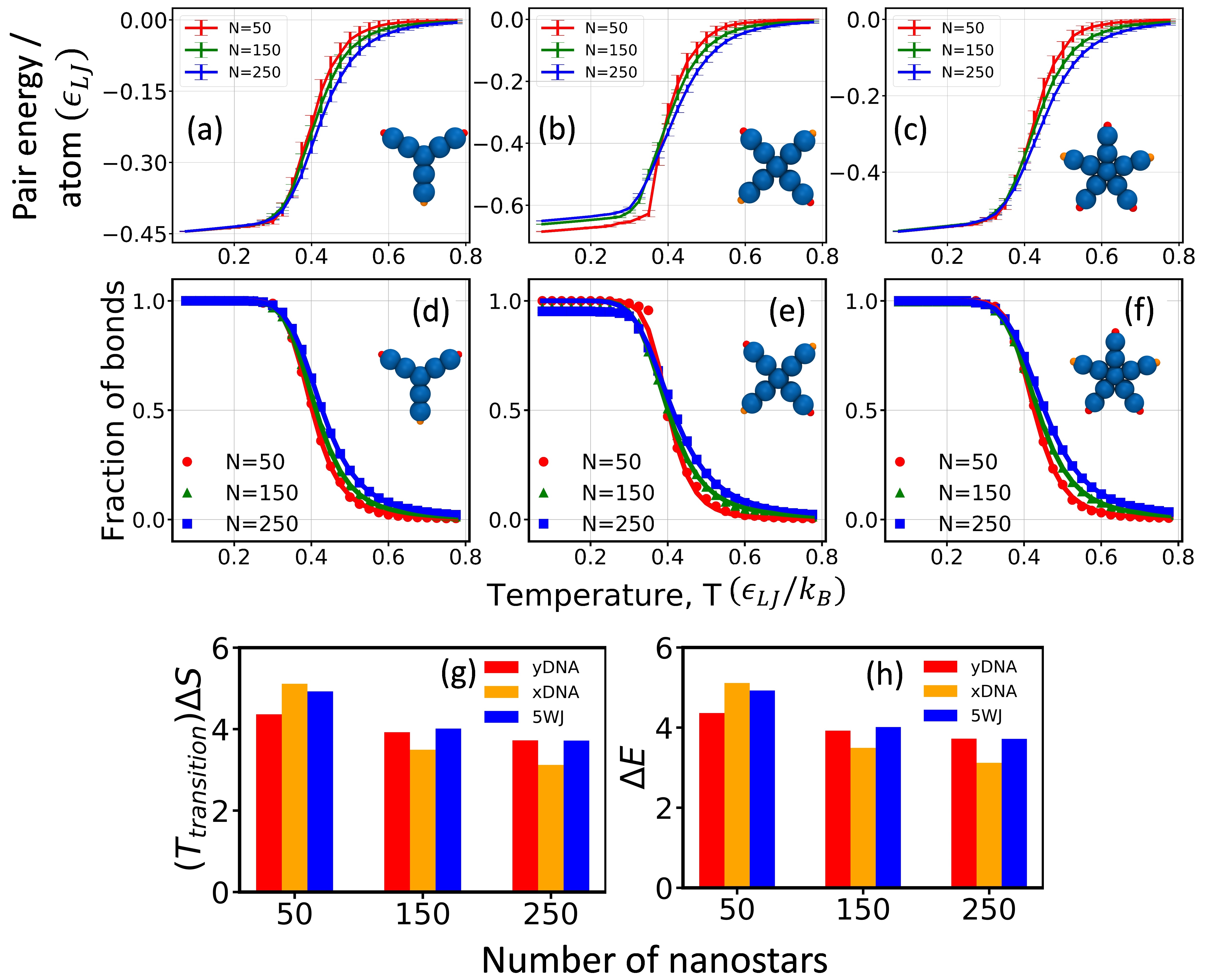}
 \caption{Potential energy (PE) per atom of the nanostar mixtures at different temperature and densities. The systems consist of (a) Y-shaped (b) X-shaped (c) 5WJ nanostars. Fraction of intact bonds at different temperature and densities. The systems consist of (d) Y-shaped (e) X-shaped (f) 5WJ nanostars. (g) Change in entropy and (h) change in enthalpy as the system goes from gel state to unstructured state. }
 \label{ch7_fig5}
\end{figure}

The formation of the network is mainly driven by the interaction between the patchy beads resulting in the formation of bonds. At low temperature, this bond formation in the gel phase gives rise to large gain in enthalpy. To have a quantitative understanding, we have computed the potential energy (PE) of the nanostar mixture at different temperatures. In figure \ref{ch7_fig5}, we have plotted the PE per atom for all the nanostars at different temperatures. We find  that the PE is positive when the temperature is high. PE becomes negative at low temperatures. Also, as we increase the temperature we observe a discontinuity in the PE over a small range of temperature (around $T^*  \sim 0.3$ to $0.5$ ). The nanostars undergo from a fully bonded state to a unstructured state within this temperature range. The discontinuity also implies the systems go through a phase transition as the temperature is increased or decreased. This morphological transition from a fully bonded state to an unstructured fluid can be well described by a two-state theoretical model as discussed below. \par

To define the bond formation between two nanostars or association between two nanostars, we define a geometric distance cutoff of 0.5 $\sigma_{LJ}$. Since the non-bonded PE between the patches is negative when they are within this confining cutoff distance, our definition of bonded network is reasonable. We calculate the total number of bonds in each system and  divide it by the number of nanostars present in the system to get the fraction of connected bonds. Similar to the behaviour of PE, we observe the system goes from a state where none of the nanostars are bonded to a state where all the nanostars are bonded and form a percolating cluster [figure \ref{ch7_fig5} (d)-(f)]. Several experimental and numerical results on thermo-reversible gel transition shows a similar behaviour\cite{chap7_sim1,chap7_sim2,chap7_sim3,chap7_sim4,chap7_sim5,chap7_largo2007self,chap7_expt9}. Also, we find that this behaviour is almost independent of the densities of the system. This sharp phase transition has a significant contribution of free energy for nanostar association, which has primarily two part -- entropic and energetic contribution. We try to estimate the  free energy change due to the formation of bonds using a simple two-state model. According to the model, the fraction of bonds, $f$ can be written as, 
\begin{equation}
f=1-\cfrac{1}{1-exp\left[\left(\Delta E - T \Delta S \right)/T\right]}    \label{ch7_7}
\end{equation}

Where, $\Delta E$ is change in enthalpy and $\Delta S$ is change in entropy upon the phase transition. We fitted the equation \ref{ch7_7} for the fraction of bonds with  $\Delta E$ and $\Delta S$ as the fitting parameter [figure \ref{ch7_fig5} (d)-(f)]. We find that the entropic change is exactly same to the enthalpic change upon gel to unstructured fluid phase transition [figure \ref{ch7_fig5} (g), (h) and table \ref{ch7_table2}]. The entropic change originates from localization of the patchy sites in a small confining region when bonded compared to the unbonded state where nanostars have full orientation and translation freedom. The contribution of enthalpy change coming from the bond formation between the patchy beads in the gel phase. This enthalpy change due to formation of bonds is comparable to the depth of attractive WCA potential of the patchy beads which in our model equivalent to 4.5$\epsilon$. The entropic loss in the gel phase is compensated by the enthalpic gain. This large enthalpic gain due to patchy bead interaction is primarily responsible for this phase transition from unstructured liquid to gel phase. Next, we try to estimate how much translational entropy contributes in this phase transition. The total pair distribution function ($g(r,\theta)$) of a system can be decomposed in two parts- radial pair distribution function ($g(r)$) and orientational pair distribution function ($g(\theta|r)$),
\begin{equation}
g(r,\theta) = g(r)g(\theta|r)  \label{ch7_8}
\end{equation}

\begin{figure}[htbp]
 \centering
 \includegraphics[width=\linewidth,keepaspectratio=true]{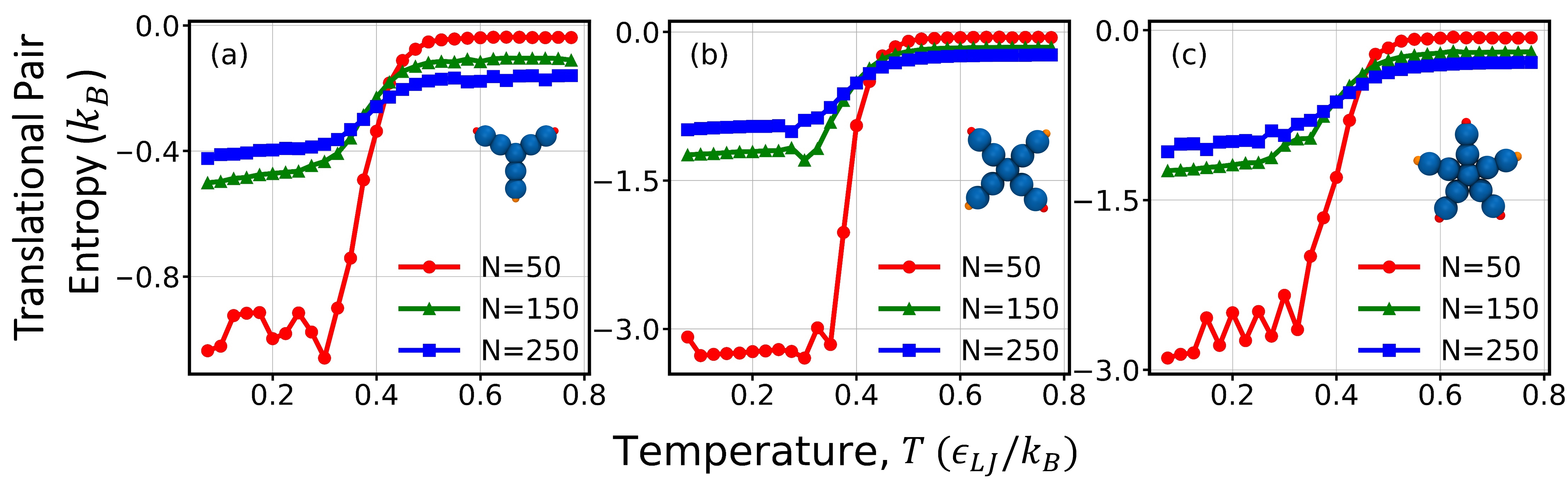}
 \caption{Translation pair entropy of different nanostar systems at different temperature and densities. The systems consist of (a) Y-shaped (b) X-shaped (c) 5WJ nanostars.}
 \label{ch7_fig6}
\end{figure}

For a complex molecule like DNA nanostar,  the orientational pair distribution function and consequently the orientational pair entropy is difficult to estimate. However, the translational pair entropy can be estimated from radial distribution function by the following relation\cite{chap7_pair},
\begin{equation}
S^{tr}_{pair}=-\frac{1}{2} \rho \int \{g(r)ln[g(r)]-g(r)+1 \}r^2 dr \label{ch7_9}
\end{equation}

The calculated translational pair entropy at different temperatures show a similar behaviour as potential energy [figure \ref{ch7_fig6}]. When the nanostars are bonded and forms gel, they loses many degrees of freedom and the translational pair entropy of the system becomes significantly low [table \ref{ch7_table2}]. But, as the temperature increases and fraction of connected bond reduces, the system gains more degrees of freedom and translational pair entropy increases [table \ref{ch7_table2}]. Also, we observe that with increasing density, the translational pair entropy for each nanostar system decreases. This is due to the fact that, when the density for a system increases the available free volume fraction decreases. The reduction of free volume fraction in turn reduces the translation motion of each nanostar. Therefore, the  translational pair entropy of the system got decreased with increased density. The qualitative characteristic of the translational pair entropy change with temperature is similar for all the systems. \par
\begin{figure}[htbp]
 \centering
 \includegraphics[width=\linewidth,keepaspectratio=true]{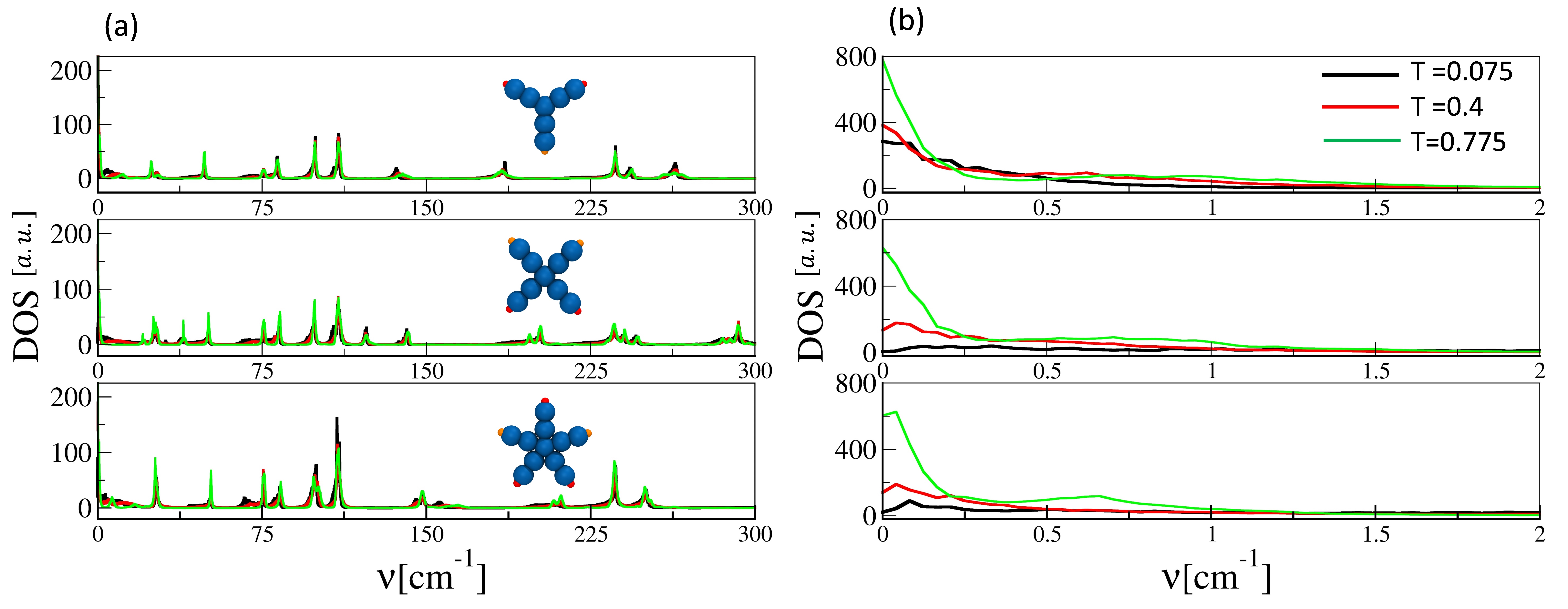}
 \caption{Density of states (DoS) of different nanostar systems at three different temperature. (a) Total DoS for all frequency regime and  (b) DoS at low frequency regime.}
 \label{ch7_dos}
\end{figure}

It is now apparent that the system losses entropy as it goes to gel phase. In order to evaluate the absolute entropy of the system, we have employed the robust Two-Phase thermodynamic (2PT) model developed by Lin et al\cite{chap7_2pt1,chap7_2pt2,chap7_2pt3}. The 2PT method built upon the simple hypothesis that the density of states(DoS) can be decomposed into solid-like (oscillatory) and gas-like (diffusive) components. First, for any poly-atomic molecule the total DoS can be  decomposed into translational, rotational and vibrational part,
\begin{equation}
\Psi^{tot}(\nu) = \Psi^{trn}(\nu) +\Psi^{rot}(\nu) +\Psi^{vib}(\nu)
 \label{ch7_10}
\end{equation}

The DoS is computed from the Fourier transform of the velocity autocorrelation function. The total DoS of each nanostar system in different frequency regime is plotted in figure \ref{ch7_dos} (a) and (b). For all the systems, we find the DoS has a pronounced peak at low frequency regime whose height increases with increasing temperature. At high temperature, the DoS exponentially decays which indicates the structures are more gas like. However, at low temperature the DOS first increases then decreases. This Dos at low temperature regime is similar to that of DoS of liquid systems, which indicates that some structural ordering is emerging in the system. Also, at high frequency range, the DoS has several peaks. These peaks corresponds to different translational, orientational stretching and vibrational modes of the nanostars. To validate that the DoS correctly captures the dynamics of the system, we computed the diffusion constant of the system from the zero frequency DoS using the following relation, $S(0)=\frac{12mND}{K_BT}$, where $D$ is the diffusion constant. We observe the diffusion constant values are very close to the diffusion constant obtained from the slope of mean-squared displacement versus time data. More details about the diffusion constant calculation are in the next section.

\begin{figure}[htbp]
 \centering
 \includegraphics[width=\linewidth,keepaspectratio=true]{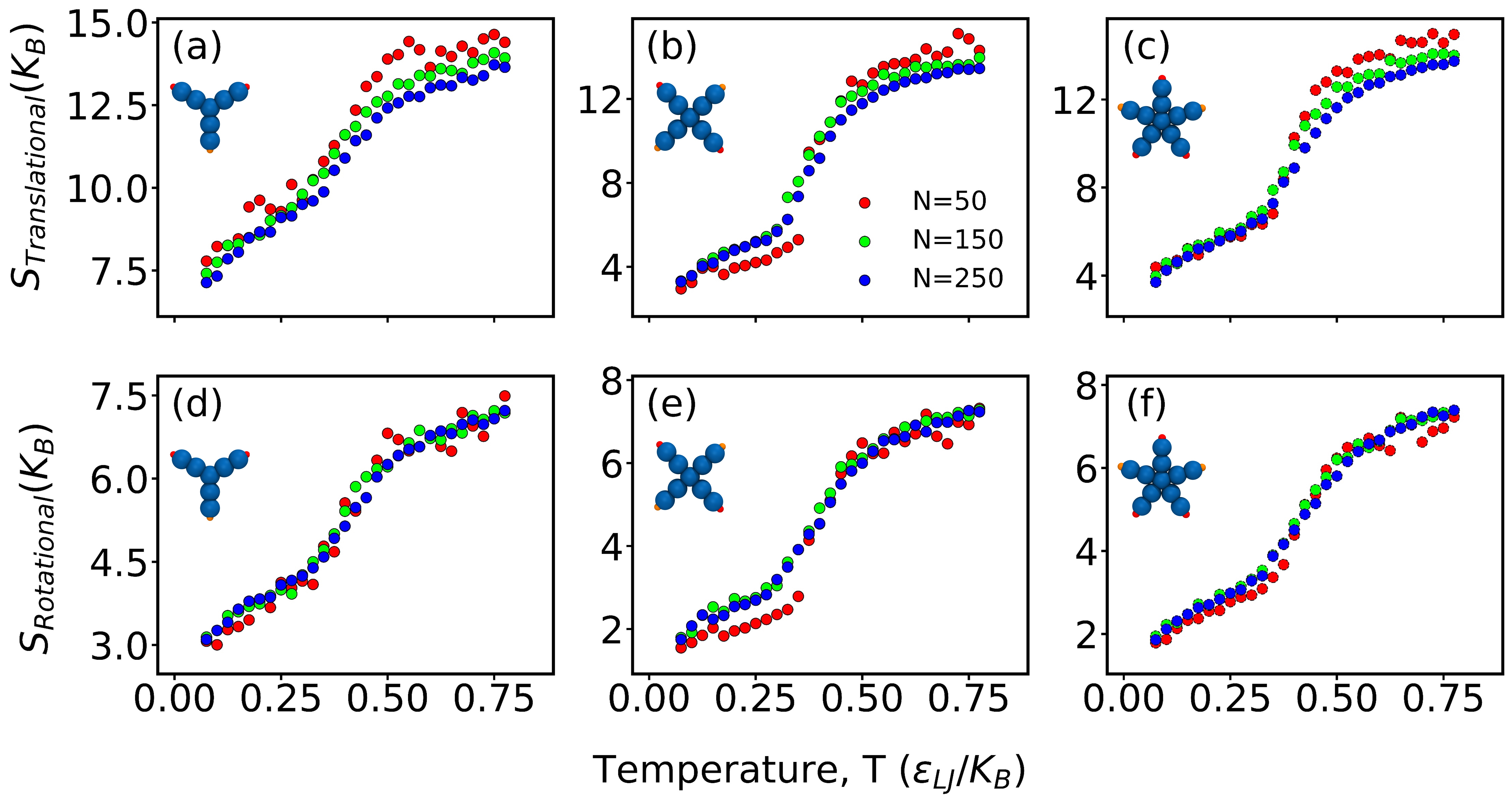}
 \caption{Absolute translational and rotational pair entropy per nanostar for different systems at different temperature and densities. Translational entropy of (a) Y-DNA, (b) X-DNA, (c) 5WJ system. Rotational entropy of (d) Y-DNA, (e) X-DNA, (f) 5WJ system.}
 \label{ch7_fig7} 
\end{figure}

In 2PT model each part of the DoS is decomposed into solid-like and gas-like contribution. 

\begin{equation}
\Psi^{\alpha}(\nu) =\Psi^{\alpha}_{solid}(\nu) +\Psi^{\alpha}_{gas}(\nu) ; \alpha = trn/rot/vib
 \label{ch7_11}
\end{equation}

Then the entropy or any thermodynamics quantity is determined from the DoS by introducing the weighting factor $W_{solid}$ and $W_{gas}$,
\begin{equation}
S^{\alpha}_{DOS} = k\left[\int_{0}^{\infty}d\nu \Psi^{\alpha}_{solid}(\nu) W_{solid} + \int_{0}^{\infty}d\nu \Psi^{\alpha}_{gas}(\nu) W_{gas} \right] ; \alpha = trn/rot/vib
 \label{ch7_13}
\end{equation}
Using the 2PT method, we next evaluated the absolute translational entropy ($S_{translational}$) and rotational entropy ($S_{rotational}$) for different nanostar systems [figure \ref{ch7_fig7}]. First, the LJ units of are converted into real unit using Argon interaction parameters and used in the 2PT code. Once entropy values are obtained using the 2PT code, results are again mapped to the LJ units. The computed $S_{translational}$ and $S_{rotational}$ show a similar behaviour to the other thermodynamics parameters like PE and transitional pair entropy. As the temperature increases, both $S_{translational}$ and $S_{rotational}$  increase with temperature and at around temperature $ 0.4\ \epsilon_{LJ} / k_{B} $, there is an abrupt jump in the entropy.  Also, the systems loss more translational entropy  compared to the rotational entropy in the gel phase [table \ref{ch7_table2}]. Another interesting fact about the gel system is that unlike liquid they loose huge rotational entropy as the temperature is decreased. And this specific property makes it distinguishable from conventional liquid phase.  

\begin{table}[htbp]
\centering
\caption{Entropy change between unstructured fluid and gel phase, calculated using different methods.$^*$}
\label{ch7_table2}
\newcolumntype{Y}{>{\centering\arraybackslash}X}
\begin{tabularx}{\textwidth}{@{}YYYYYYYY@{}}
\\
\hline 
\hline
System & Number of nanostars & Gel-to-sol transition temperature & Enthalpy change from two-state Model & Entropy change from two-state Model & Translational pair entropy change & Absolute translational  entropy change & Absolute rotational  entropy change  \\
\hline
Y-DNA  & 50       & 0.40 &     4.36     & 10.84 & 0.92                             & 5.95           &  3.89    \\
       & 150      & 0.41 &     3.92     & 9.51  & 0.38                             & 5.78           &  3.64    \\
       & 250      & 0.43 &     3.72     & 8.69  & 0.25                             & 5.70           &  3.62      \\
\hline
X-DNA  & 50       & 0.40 &      5.11    & 12.73 & 3.16                             & 10.94           &  5.08    \\
       & 150      & 0.40 &      3.49    & 8.72  & 1.07                             &  9.64           &   4.97   \\
       & 250      & 0.41 &      3.12    & 7.57  & 0.73                             &  9.42           &   4.97   \\
\hline
5WJ    & 50       & 0.43 &      4.92    & 11.55 & 2.72                             & 10.02           & 4.85     \\
       & 150      & 0.44 &      4.01    & 9.20  & 1.03                             & 9.23            &  4.92    \\
       & 250      & 0.45 &      3.71    & 8.21  & 0.74                             & 9.01            &   4.97   \\
\hline
\\

\end{tabularx}

$^*$Entropy difference between the gel phase and fluid phase for the three former methods is evaluated by taking the difference of average entropy between the two states. Average entropy is calculated by averaging extreme five state points of entropy vs temperature graphs.  \\
\hrulefill
\end{table}

\subsection{\NoCaseChange{Dynamics}}
\begin{figure}[htbp]
 \centering
 \includegraphics[width=6.2in,keepaspectratio=true]{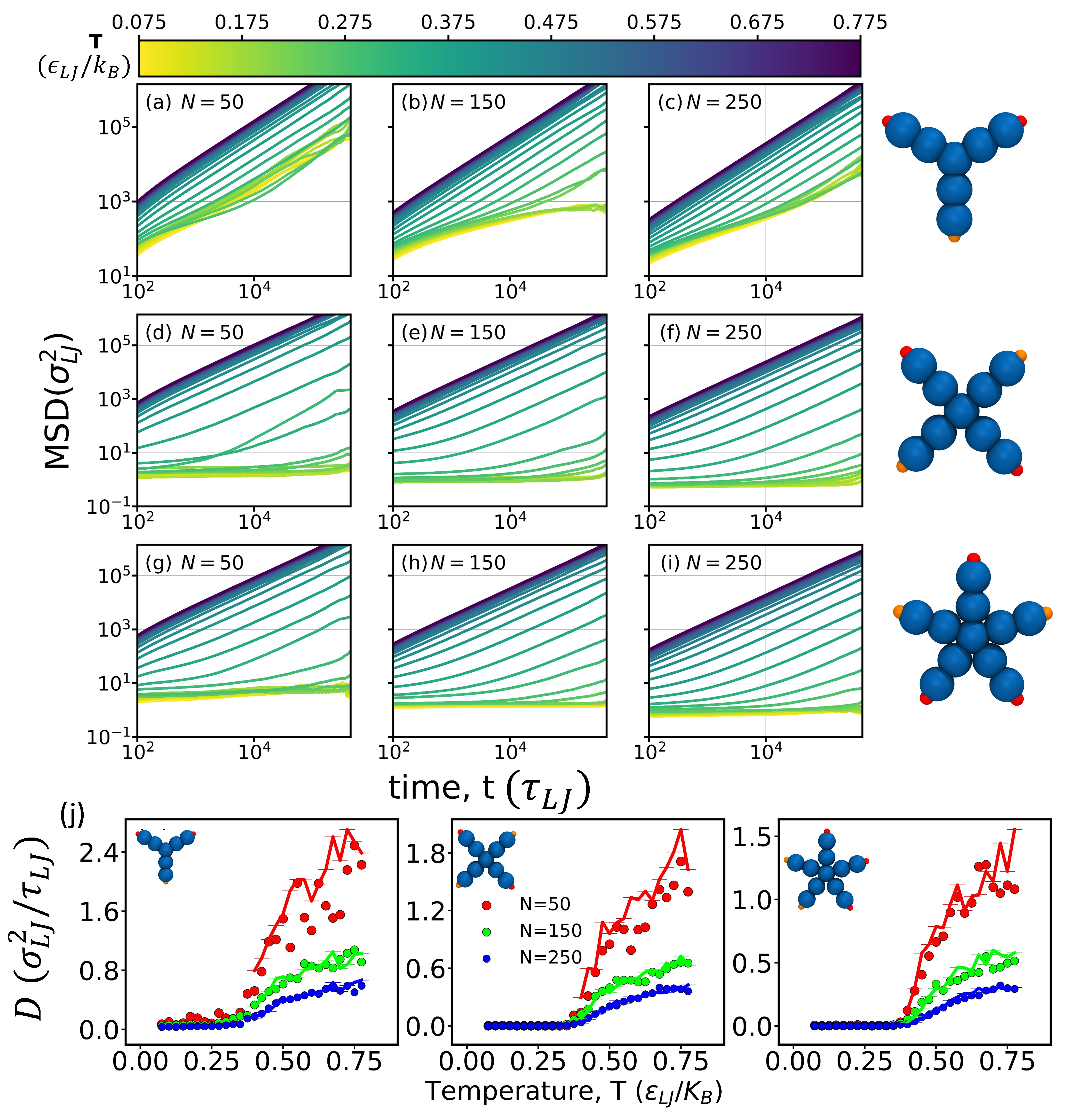}
 \caption{Mean squared displacement (MSD) of different nanostar systems at different temperature. (a)-(c) MSD of Y-shaped nanostar mixtures. The number of nanostars in the systems is (a) $N=50$, (b) $N=150$, (c) $N=250$. (d)-(f) MSD of X-shaped nanostar mixtures. The number of nanostars in the systems is (d) $N=50$, (e) $N=150$, (f) $N=250$. (g)-(i) MSD of 5WJ nanostar mixtures. The number of nanostars in the systems is (g) $N=50$, (h) $N=150$, (i) $N=250$. (j) Diffusion constant of different nanostar systems at different temperature and densities. Points are calculated using velocity autocorrelation and line corresponds to diffusion constant from the slope of MSD vs time graph.}
 \label{ch7_fig8}
\end{figure}

To better understand the structure and thermodynamics of the gelation with the dynamical arrest of the system, we next computed the mean square displacement (MSD) as a function of time for different temperatures and densities. The MSD of the central bead of each nanostar is evaluated using the following relation, 
\begin{equation}
MSD = \langle r^2 \rangle = \cfrac{1}{N} \sum_{i=1}^{N} \langle [r_i (t+t^{'}) - r_i(t)]^{2} \rangle \label{ch7_15}
\end{equation}
where $N$ is total number of beads, $t$ is time difference, $t^{'}$ is time origin and the angular bracket denotes an average over all the time origins. The calculated MSD of different systems is plotted in figure \ref{ch7_fig8}.  Diffusion constant is calculated from the linear region of the MSD using the following relation, 
\begin{equation}
MSD = \langle r^2 \rangle = 6 D t  \label{ch7_16}
\end{equation}
where D is the diffusion constant of the system. At high temperature, the MSD of the system increases linearly with time. Also, as the temperature decreases and the nanostars makes bonds with one another, the diffusion constant also decreases [figure \ref{ch7_fig8}]. Moreover, when the system forms complete percolating network and goes to gel phase the Einstein relation of diffusion constant i.e., equation \ref{ch7_16} no longer holds. So, we are restricted to calculate the diffusion constant up to temperature, $T^* = 0.4$. Also, we have calculated the diffusion constant from the density of states at zero frequency from the 2PT method, 
\begin{equation}
S_{DOS} = \frac{2}{K_B T} \int_{-\infty}^{\infty} C(t) dt = \frac{12mND}{K_B T}
\end{equation}
Where C(t) is the mass weighted sum of the velocity autocorrelation function.The diffusion constant calculated using 2PT method is slightly higher compared to the diffusion constant calculated from MSD. As in MSD calculation, we only considered the central bead which removes any kind rotational motion. However, in 2PT method, it considers translational, rotational and vibrational motion of each nanostar while calculating the velocity autocorrelation function. Thus, the diffusion constant in MSD is slightly underestimated. Also, we find that with increasing density, the value of D decreases for every system. This is expected as with increasing density, the accessible phase space volume of the nanostars decreases. Another interesting fact is that the diffusion constant for any particular density is highest for Y-DNA and lowest for 5WJ nanostars while the diffusion constant of X-DNA lies in between them. This can also be explained from the  volume fraction of the each system. The volume fraction of 5WJ is highest compared to the other two nanostar systems which reduces the available phase space volume to diffuse and this is reflected in the overall diffusion of the system.

\section{\NoCaseChange{Conclusion}}
In summary, we have presented a bead-spring model of DNA nanostar to study the structure and governing dynamics and thermodynamics of their gelation. This simple model captures the essential qualitative feature of DNA gelation for a wide range of density. Using RDF calculation, we have shown that all the systems transform from unstructured fluid to a gel like phase upon temperature lowering. The amorphous nature of the gel phase was confirmed using structure factor calculation. The phase transition from arrested gel phase to a unstructured liquid phase has been modelled using two-state theoretical model. We find loss of configuration and translational entropy is mainly giving rise to gel phase at low temperature. There is a competition between the entropic loss  and the enthalpic gain in the gel phase. The absolute translational and rotational entropy of the system which is evaluated using two-phase thermodynamic model, also substantiate the above speculation.  The slowing down of the dynamics upon approaching the transition temperature, demonstrates the phase transition to the gel phase. The present work offers a thorough understanding of DNA hydrogel formation for different complex nanostars. Also, in this work, we have made an effort to evaluate the absolute entropy of a model gel system and tried to explain the possible origin of gel formation from a thermodynamic perspective. Further, this model can be implemented to study the gel formation of similar other chemical species where bond formation is mediated by hydrogen bonding by simply tuning the force-field parameters. We believe that the study will be helpful in understanding the underlying physics of gel formation of hyper-branched DNA like nanostars for various bio- and nano-technological applications.

\section{\NoCaseChange{Acknowledgement}}
SN thanks IISc for RA fellowship. DB thanks SERB-DST for Ramanujan Fellowship; DBT-EMT, Gujcost-DST, GSBTM for Research grant and IITGN for initial startup support. We thank IISc Bangalore for the computational support. SN thanks Jayeeta Chattopadhyay for critically reading the manuscript.

\section{\NoCaseChange{Conflict of  interest}}
The authors have no conflict of interest to declare.

\bibliographystyle{apsrev4-2}
\bibliography{main}
\end{document}